\documentclass{PoS}

\title{Chiral transition temperature and aspects of deconfinement 
in {2+1}~flavor QCD with the HISQ/tree action}

\ShortTitle{Chiral transition temperature and aspects of
deconfinement in 2+1 flavor QCD}

\author{\speaker{A.~Bazavov} (for HotQCD 
Collaboration)\thanks{
HotQCD Collaboration members are:
A.~Bazavov,
T.~Bhattacharya,
M.~Cheng,
N.H.~Christ,
C.~DeTar,
H.-T.~Ding,
S.~Gottlieb,
R.~Gupta,
P.~Hegde,
U.M.~Heller,
C.~Jung,
F.~Karsch,
E.~Laermann,
L.~Levkova,
R.D.~Mawhinney,
S.~Mukherjee,
P.~Petreczky,
D.~Renfrew,
C.~Schmidt,
R.A.~Soltz,
W.~Soeldner,
R.~Sugar,
D.~Toussaint,
P.~Vranas}\\
  Physics Department, Brookhaven National Laboratory, Upton, NY 11973}

\abstract{
We present results on the chiral transition temperature $T_c$ 
in 2+1 flavor QCD extrapolated to the continuum limit and
the physical light quark mass. The extrapolations are based 
on the data from simulations on lattices with temporal
extent $N_{\tau}=6$, $8$ and $12$ with the HISQ/tree and
$N_\tau=8$ and 12 with the asqtad action.
The chiral transition is analyzed in terms of universal 
$O(N)$ scaling functions.
After performing simultaneous asqtad and HISQ/tree continuum
extrapolation the chiral transition temperature is 
$T_c=154 \pm  9$ MeV. We also discuss the deconfinement
aspects of the transition in terms of the renormalized
Polyakov loop, fluctuations and correlations of several
conserved charges and the trace anomaly.
}

\FullConference{XXIX International Symposium on Lattice Field Theory \\
                 July 10 -- 16 2011\\
                 Squaw Valley, Lake Tahoe, California}

\begin{document}

\section{Introduction}
\label{sec: intro}
In these proceedings we follow up on the line of work on 2+1 flavor
QCD thermodynamics with improved staggered fermions by the HotQCD
collaboration. The setup of simulations and some preliminary 
results have been reported earlier, e.g.
\cite{Jamaica,lat10,soeldner_lat10,prasad_panic11}
and the continuum extrapolation for chiral $T_c$ at the physical
quark masses is presented in \cite{hotqcd2}.
Full data set includes several lines of constant physics
down to the light quark mass $m_l=m_s/20$ for asqtad and
$m_l=m_s/40$ for HISQ/tree\footnote{The lightest mass $m_l=m_s/40$
is used for the analysis of the chiral condensate and 
susceptibility~\cite{future},
while all other quantities are calculated at $m_l=m_s/20$ for
HISQ/tree.}.
The lattice spacings cover the range of temperatures
$T=130-440$ MeV with $N_\tau=6$, 8 and 12 for HISQ/tree and
$T=148-304$ MeV with $N_\tau=8$ and 12 for asqtad.

Performing the continuum limit requires control over cutoff effects.
In improved staggered discretization schemes the leading $O(a^2)$
errors at low temperature (coarse lattices) originate from 
violations of taste symmetry, that distort the hadron spectrum.
For this reason it is important to perform simulations on fine
enough lattices (large $N_\tau$ in finite-temperature setup)
and/or use actions with the smallest discretization effects.
Analysis of the discretization effects for asqtad and HISQ/tree
used in this study is presented in Ref.~\cite{lat10}.

\section{Chiral transition}
For vanishing light quark masses there is a chiral phase transition which is
expected to be of second order and in the $O(4)$ universality class~\cite{rob_o4}.
However, universal scaling allows to define pseudo-critical temperatures for
the chiral transition even for non-zero light quark masses, provided they are small
enough. For staggered fermions that preserve only a part of the chiral symmetry
there is a complication: in the chiral limit at finite lattice spacing the relevant
universality class is $O(2)$ rather than $O(4)$. Fortunately,
in the numerical analysis the differences between $O(2)$ and $O(4)$ universality classes
are small so when referring to scaling we will use the term $O(N)$ scaling. 
Previous studies with the p4 action provided evidence for $O(N)$ scaling \cite{rbcbi09,rbcbi10}.
Similar analysis for the asqtad and HISQ/tree action establishing
if the $O(N)$ scaling is applicable is performed in~\cite{hotqcd2}
and explained below. 

The order parameter for the chiral transition is the chiral condensate
\begin{equation}
M_b \equiv \frac{m_s \langle \bar{\psi}\psi \rangle_l}{T^4} \; .
\label{order}
\end{equation}
Its temperature and quark mass dependence near the critical temperature
can be parametrized by a universal scaling function $f_G$ and a regular
function $f_{M,reg}$ that describes corrections to scaling:
\begin{equation}
M_b(T,m_l,m_s) = h^{1/\delta} f_G(t/h^{1/\beta\delta}) + f_{M,reg}(T,H),
\,\,\,\,
t = \frac{1}{t_0} \left( \frac{T-T_c^0}{T_c^0} \right),\,\,\,\,
h= \frac{1}{h_0} H,\,\,\,\,H=\frac{m_l}{m_s}
\label{order_scaling}
\end{equation}
and $T_c^0$ is the critical temperature in the chiral limit.
The pseudo-critical temperature can be defined as the peak position of
the chiral susceptibility
\begin{equation}
\chi_{m,l} =  \frac{\partial}{\partial m_l}\langle \bar{\psi}\psi \rangle_l
\label{suscept}
\end{equation}
whose scaling behavior is also described by $f_G$ and $f_{M,reg}$ as
\begin{equation}
\frac{\chi_{m,l}}{T^2} = \frac{T^2}{m_s^2} 
\left( \frac{1}{h_0}
h^{1/\delta -1} f_\chi(z) + \frac{\partial f_{M,reg}(T,H)}{\partial H} 
\right),\,\,\,\,
f_{\chi}(z)=\frac{1}{\delta} [f_G(z)-\frac{z}{\beta} f_G'(z)],
\,\,\,\,z=\frac{t}{h^{1/\beta\delta}}.
\label{chiralsuscept}
\end{equation}

The singular function $f_G$ is well studied in spin models and has been parametrized
for $O(2)$ and $O(4)$ groups. For the regular part we consider leading-order (linear)
dependence in $H$ and quadratic in $T$:
\begin{equation}
f_{M,reg}(T,H) = \left( a_0 + a_1 \frac{T-T_c^0}{T_c^0} + 
a_2 \left(\frac{T-T_c^0}{T_c^0} \right)^2 \right) H.
\label{eq:freg}
\end{equation}
Then we are left with 6 parameters to be determined from fitting the data,
$T_c^0$, $t_0$, $h_0$, $a_0$, $a_1$ and $a_2$.

\begin{figure}
\includegraphics[width=0.48\textwidth]{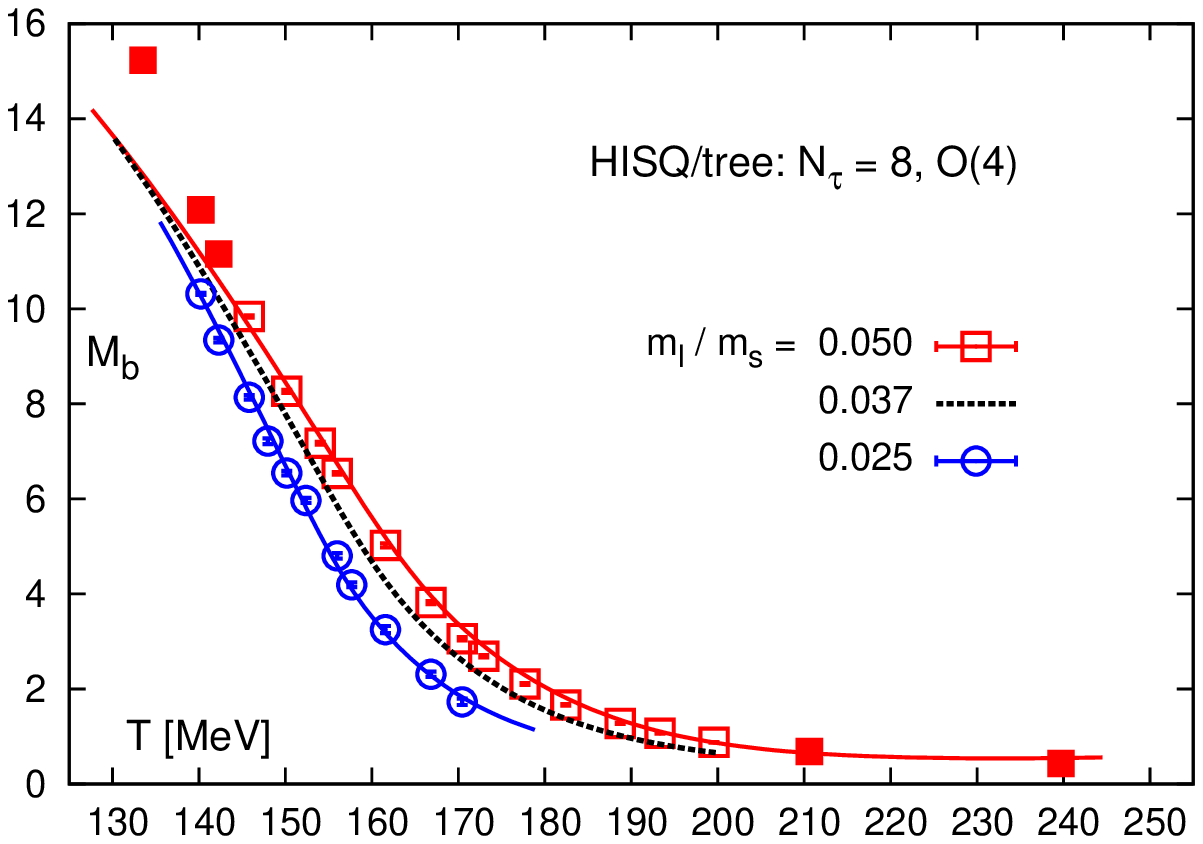}
\hfill
\includegraphics[width=0.48\textwidth]{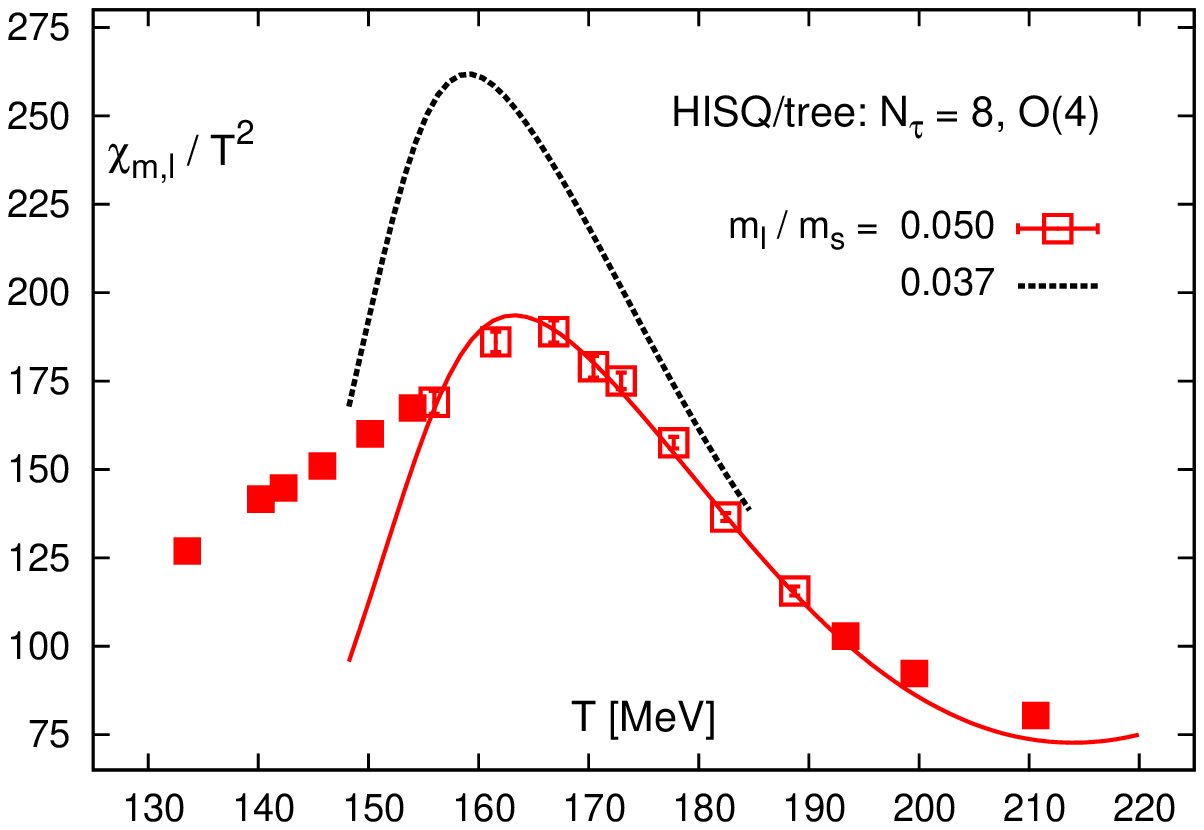}
\caption{
An example of a simultaneous fit to the chiral condensate (left)
and susceptibility (right) for HISQ/tree on $N_\tau=8$ lattices.
Open symbols indicate the range included in the fit. Dotted black
line is an extrapolation to the physical light quark mass.
}
\label{pbp_and_chi}
\end{figure}

We perform simultaneous fits to $M_b$ and $\chi_{m,l}$ for the asqtad
action on $N_\tau=8$, 12 and the HISQ/tree action on $N_\tau=6$, 8 and 12.
An example of such a fit for HISQ/tree, $N_\tau=8$ is shown in
Fig.~\ref{pbp_and_chi}.

\begin{figure}
\begin{center}
\includegraphics[width=0.48\textwidth]{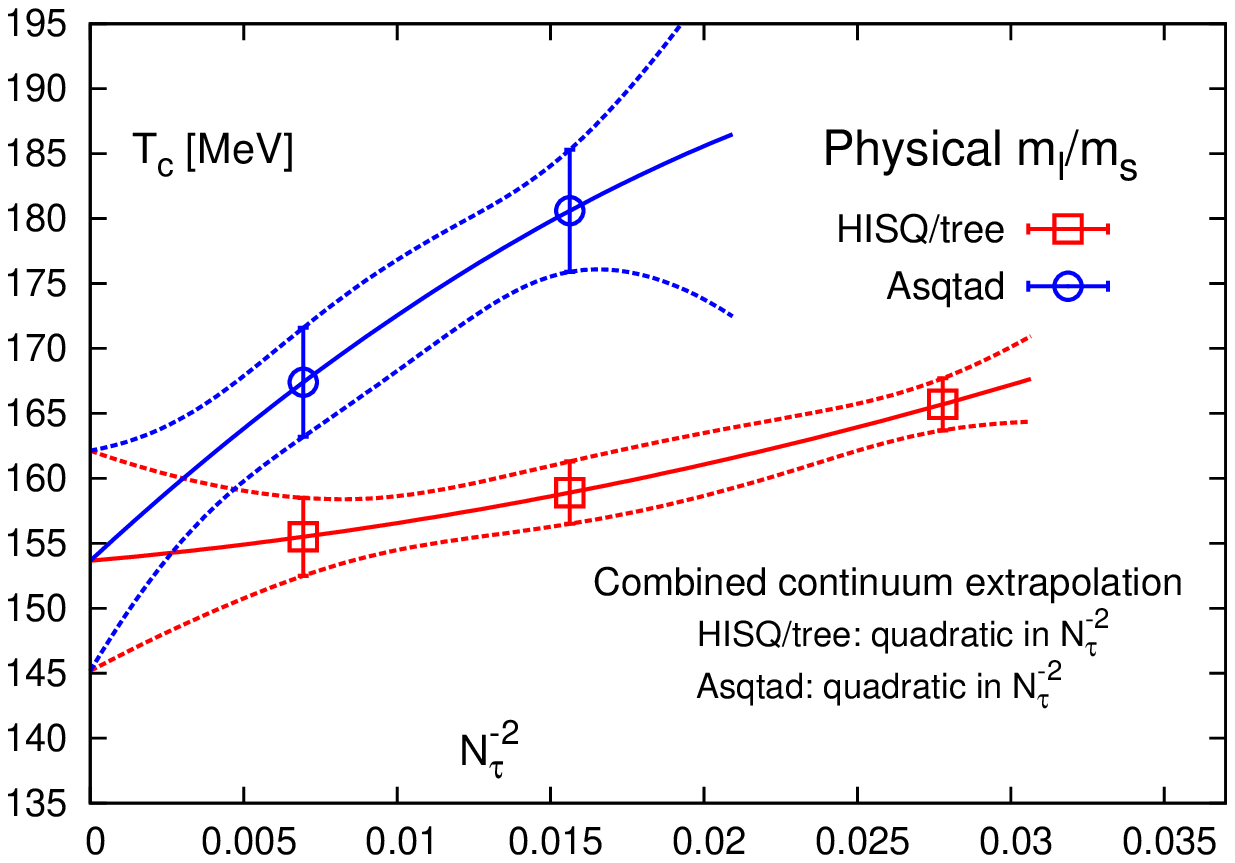}
\hfill
\includegraphics[width=0.464\textwidth]{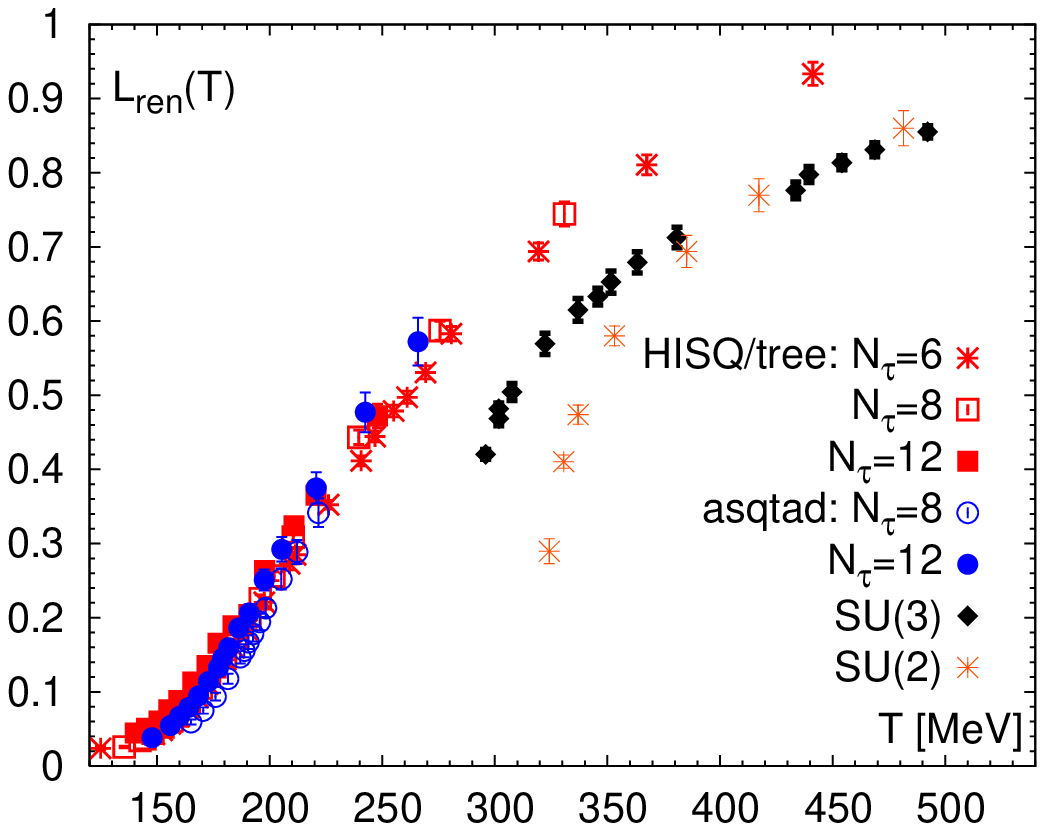}
\parbox[b]{0.48\textwidth}{
\caption{
Example of a combined fit to the asqtad and HISQ/tree data.
The fit is constrained to have a
common intercept at $1/N_\tau^2\to0$.
}
\label{tc_comb}
}
\hfill
\parbox[b]{0.464\textwidth}{
\caption{The renormalized Polyakov loop in pure gauge $SU(2)$, 
$SU(3)$ and full QCD with asqtad and HISQ/tree.}
\label{Lren_absT}
}
\end{center}
\end{figure}

Then, performing a combined  $1/N_{\tau}^2$ extrapolation of $T_c$ values obtained 
with the asqtad and HISQ/tree action as shown in Fig.~\ref{tc_comb} we obtain 
\begin{equation}
T_c=( 154 \pm 8 \pm 1)\mbox{ MeV},
\end{equation} 
where the first error is from the fit and the second is the overall
error on the lattice scale determination. The fits for asqtad and HISQ/tree are
constrained to have a common intercept. (See Ref.~\cite{hotqcd2} for more details
on the fitting procedure and analysis of systematic errors.)
To present a combined error we add the two errors giving our final
value $T_c=154\pm9$ MeV.

\section{Deconfinement aspects of the transition}

The deconfinement phenomenon in pure gauge theory is governed by
breaking of the $Z(N_c)$ symmetry. The order parameter is the 
renormalized Polyakov loop, obtained from the bare Polyakov loop as
\begin{equation}
L_{ren}(T)=z(\beta)^{N_{\tau}} L_{bare}(\beta)=
z(\beta)^{N_{\tau}} \left\langle\frac{1}{N_c}  {\rm Tr } 
\prod_{x_0=0}^{N_{\tau}-1} U_0(x_0,\vec{x})\right\rangle,
\end{equation}
where $z(\beta)=\exp(-c(\beta)/2)$. $c(\beta)$ is the additive normalization
of the static potential chosen such that it coincides with the string potential
at distance $r=1.5r_0$ with $r_0$ being the Sommer scale.

In QCD $Z(N_c)$ symmetry is explicitly broken by dynamical quarks,
therefore there is no obvious reason for the Polyakov loop to be
sensitive to the singular behavior close to the chiral limit.
Indeed, the temperature dependence of the Polyakov loop in
pure gauge theory and in QCD is quite different, as one can see 
from Fig.~\ref{Lren_absT}. Also note, that in this purely gluonic
observable there is very little sensitivity (through the sea quark
loops) to the cut-off effects coming from the fermionic sector.
While losing the status of the order parameter
in QCD, the Polyakov loop is still a good probe of screening 
of static color charges in quark-gluon plasma~\cite{okacz02,digal03}.

\begin{figure}
\includegraphics[width=0.48\textwidth]{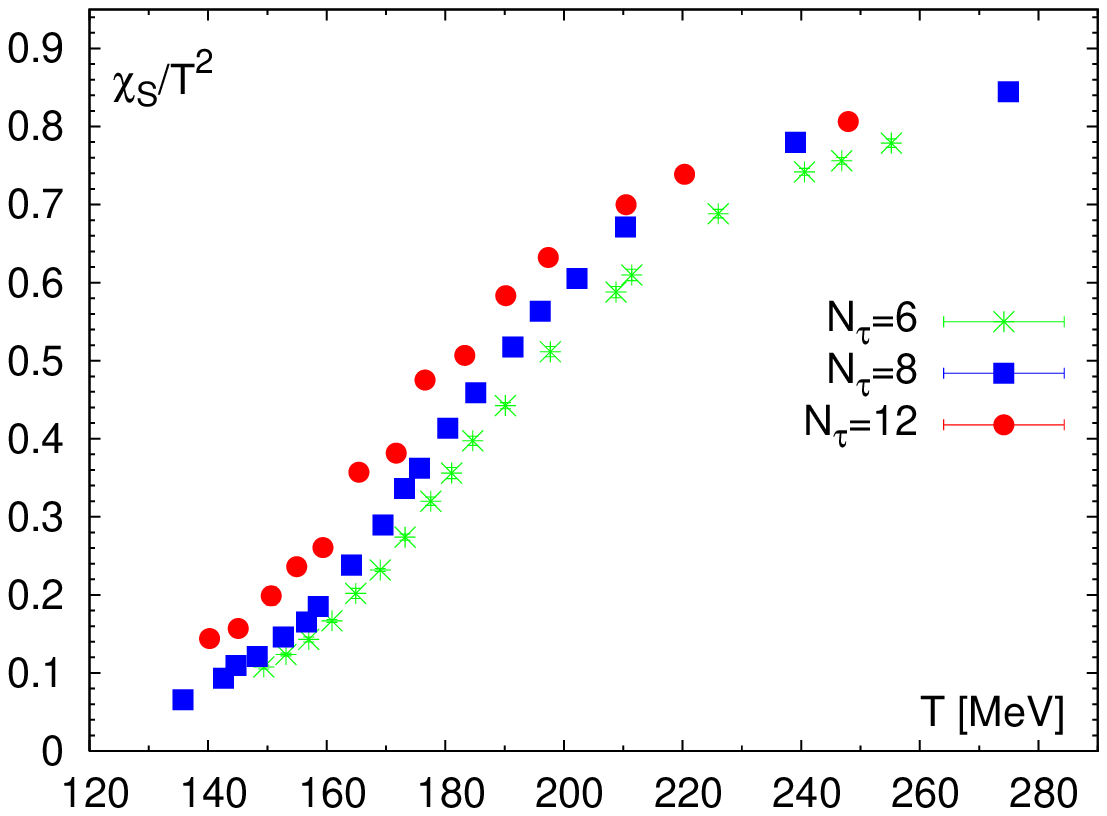}
\hfill
\includegraphics[width=0.48\textwidth]{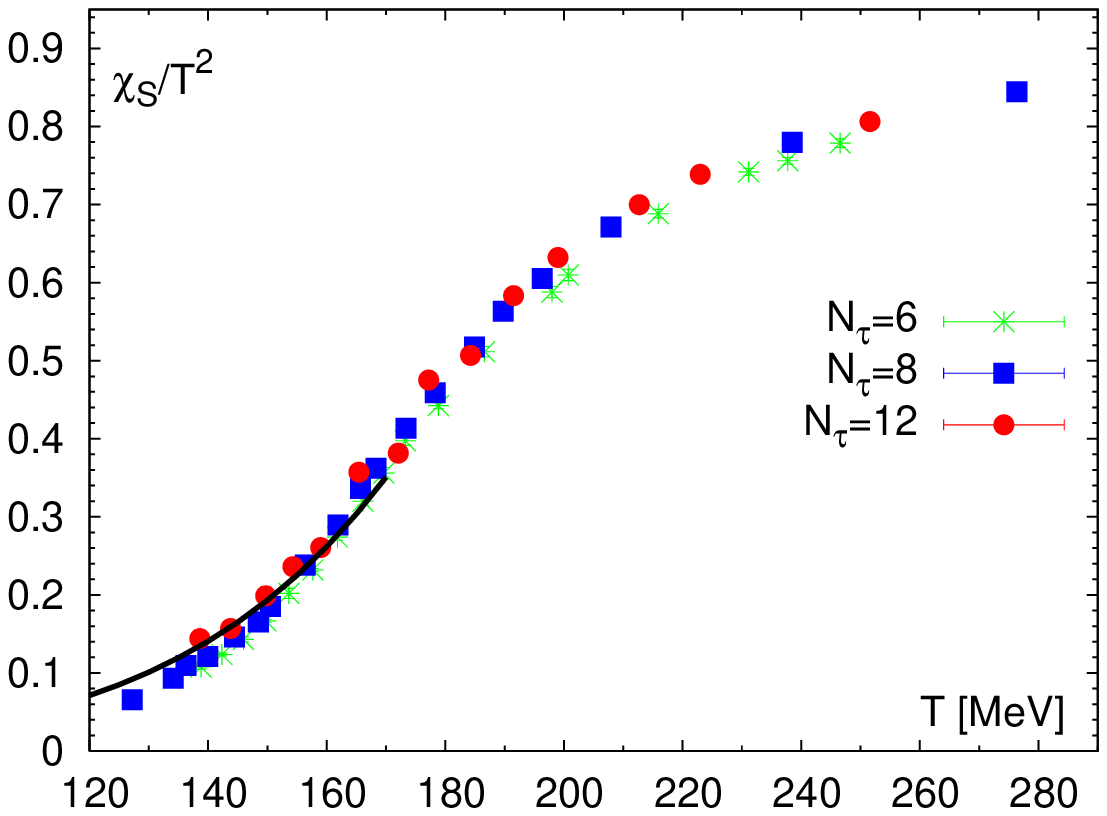}
\caption{
The strange quark number susceptibility for the HISQ/tree action
at $m_l=m_s/20$ with $r_1$ (left) and $f_K$ (right) temperature
scale. The solid curve in the right panel is from the HRG model.
}
\label{fig:qns_s}
\end{figure}

\begin{figure}
\includegraphics[width=0.48\textwidth]{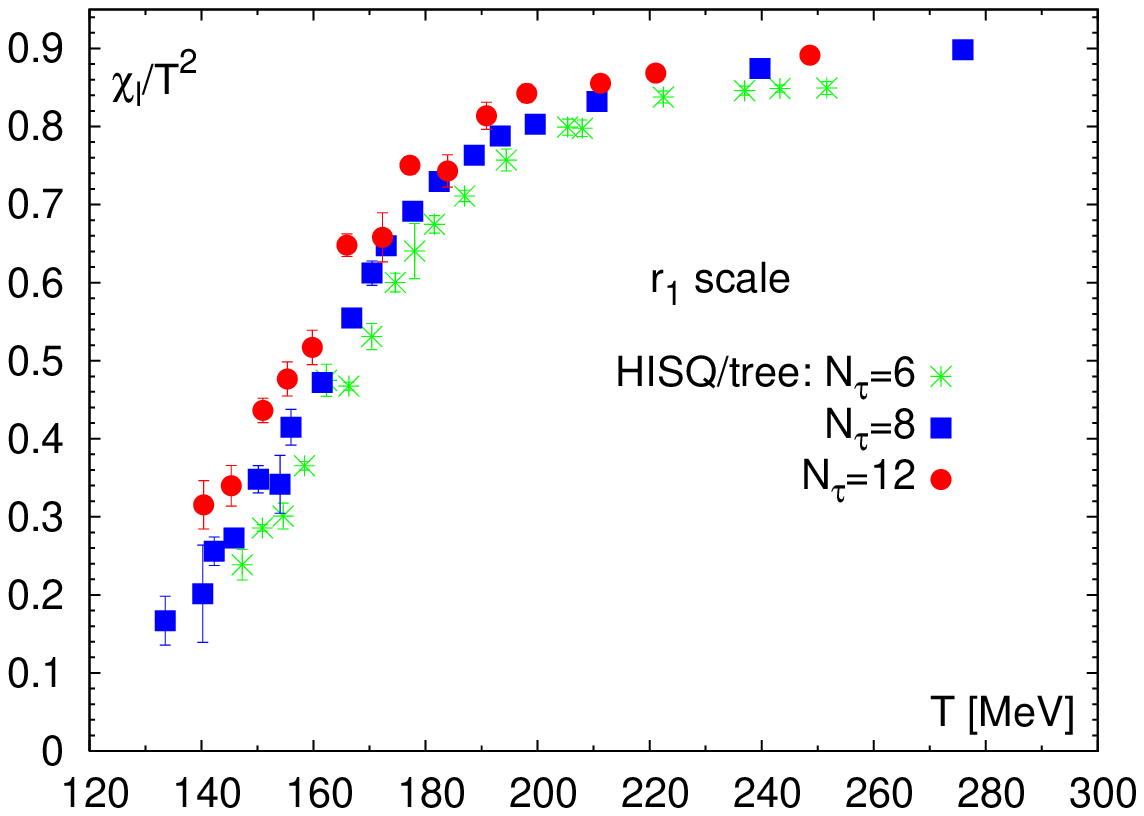}
\hfill
\includegraphics[width=0.48\textwidth]{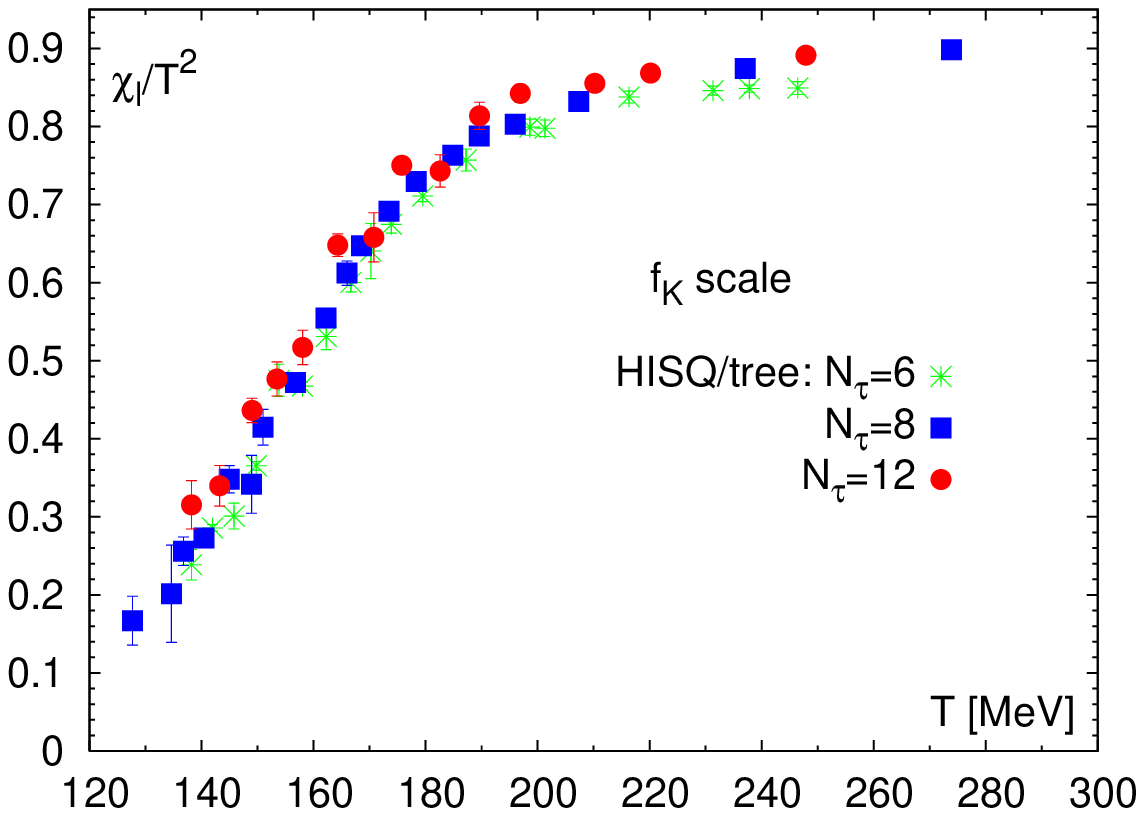}
\caption{
The light quark number susceptibility for the HISQ/tree action
at $m_l=m_s/20$ with $r_1$ (left) and $f_K$ (right) temperature
scale.
}
\label{fig:qns_l}
\end{figure}

Other probes of deconfinement are fluctuations and correlations
of various charges that can signal liberation of degrees of freedom
with quantum numbers of quarks and gluons in the high-temperature
phase. Here we consider quadratic fluctuations and correlations
of conserved charges:
\begin{eqnarray}
\frac{\chi_i(T)}{T^2}=
\left.\frac{1}{T^3 V}\frac{\partial^2 \ln Z(T,\mu_i)}{\partial (\mu_i/T)^2}
\right|_{\mu_i=0},\,\,\,\,\,
\frac{\chi_{11}^{ij}(T)}{T^2}=
\left.\frac{1}{T^3 V}\frac{\partial^2 \ln Z(T,\mu_i,\mu_j)}{\partial (\mu_i/T) 
\partial (\mu_j/T)} \right|_{\mu_i=\mu_j=0}.
\end{eqnarray}
Fluctuations are also sensitive to the singular part of the free energy
density.
The light and strange quark number susceptibilities are often
considered in connection with deconfinement. These quantities
show a rapid rise in the transition region and for the strange
quark number susceptibility the behavior is consistent with the
Hadron Resonance Gas (HRG) model, as shown in Fig.~\ref{fig:qns_s}.
For the light quark number susceptibility,
a quantity dominated by pions and therefore very sensitive to the
taste symmetry, cut-off effects are still significant, and
comparing to HRG is more subtle and requires further study, Fig.~\ref{fig:qns_l}.
In the left panels of Fig.~\ref{fig:qns_s} and \ref{fig:qns_l} 
temperature is set by Sommer scale ($r_1$ in this case).

In part cut-off effects in the observables directly related to hadrons
can be accounted for if the temperature scale is set with a hadronic 
observable. Following the insight by the Budapest-Wuppertal 
collaboration \cite{stout,stout3} we use the kaon decay constant $f_K$ for this purpose.
For the strange and light quark number susceptibility the results with
$f_K$ scale are shown in the right panels of Fig.~\ref{fig:qns_s} and
\ref{fig:qns_l}.

As one can see, for the strange quark number susceptibility the $f_K$ scale
eliminates virtually all cut-off dependence and the data from 
different lattices are hardly distinguishable. For the light quark number
susceptibility some residual cut-off dependence remains. This may be expected
for, at least, two reasons: a) $\chi_l$ is very sensitive to taste
symmetry breaking (even at our finest, $N_\tau=12$ lattice the root-mean-squared
pion mass is still about 200 MeV for the HISQ/tree action~\cite{lat10}),
b)~there is no a priori reason for the lattice spacing dependence in $\chi_l$,
$\chi_s$ to be the same as in $f_K$.

\begin{figure}
\includegraphics[width=0.48\textwidth]{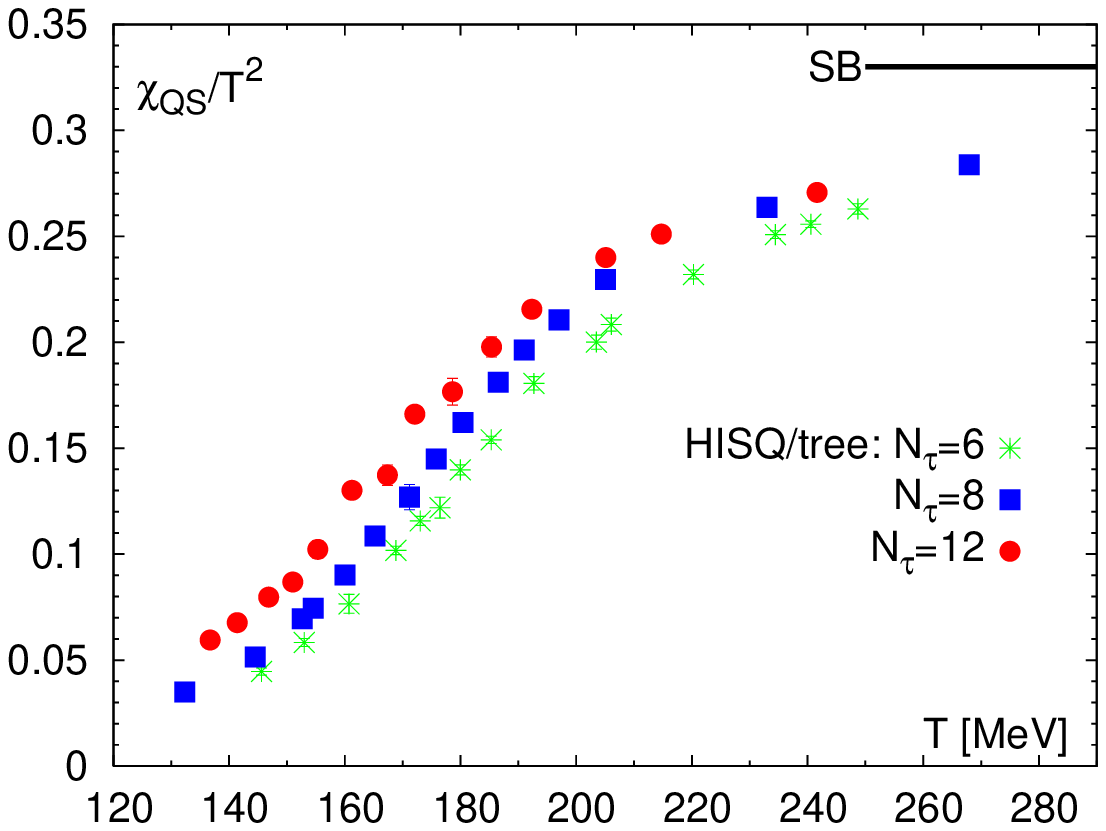}
\hfill
\includegraphics[width=0.48\textwidth]{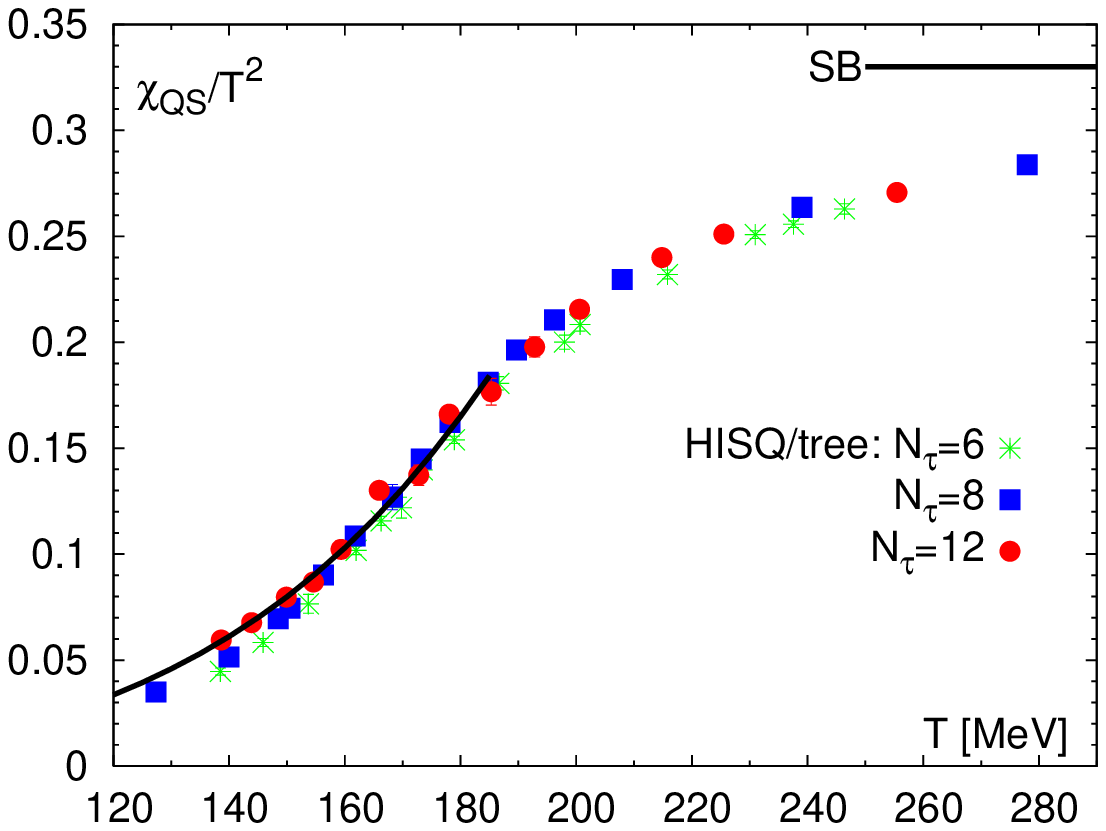}
\caption{
Electric charge--strangeness correlations with $r_1$ (left)
and $f_K$ (right) temperature scale for HISQ/tree. The solid
curve in the right panel is from the HRG model.
}
\label{fig:qns_QS}
\end{figure}

Next, we consider correlations of the electric charge and strangeness.
With the $r_1$ scale we observe substantial cut-off dependence,
which is largely reduced with $f_K$ scale,
compare left and right panel in Fig.~\ref{fig:qns_QS}. At high temperatures
these correlations are close to the value expected in the non-interacting
ideal gas (Stefan-Boltzmann limit), while at low temperatures they
are well described by HRG.

\begin{figure}
\begin{center}
\includegraphics[width=0.48\textwidth]{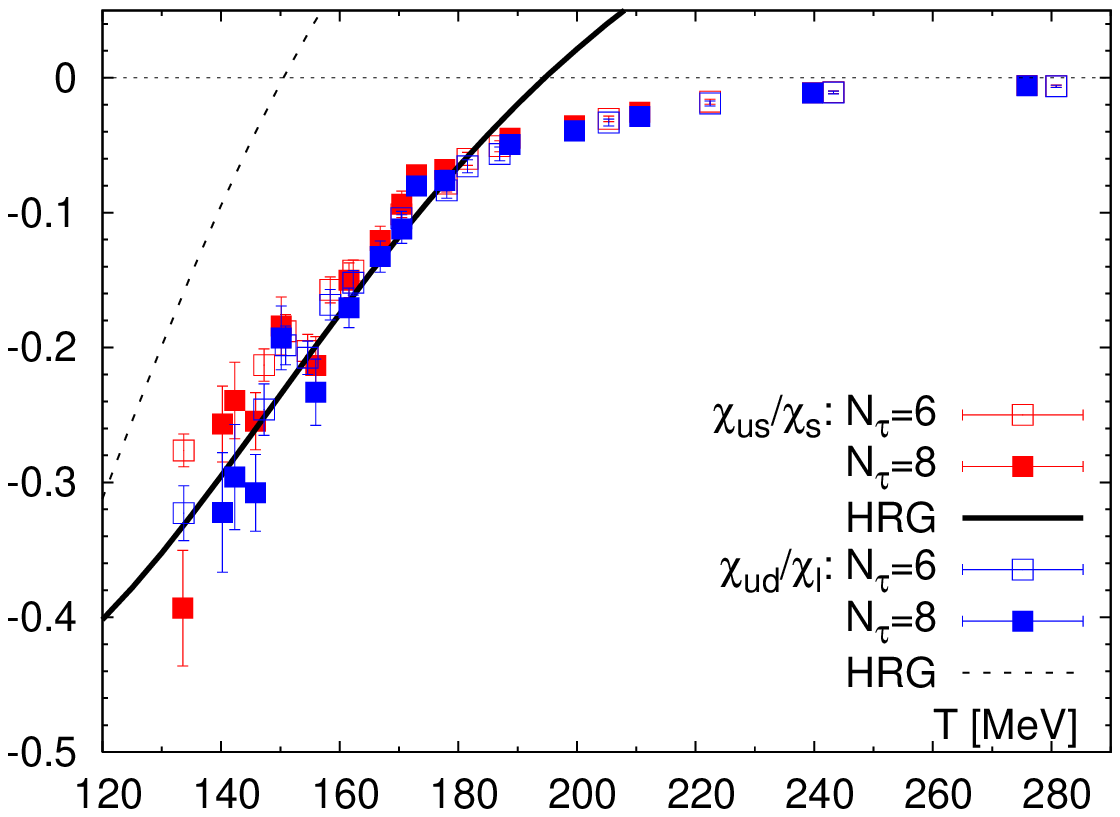}
\hfill
\includegraphics[width=0.48\textwidth]{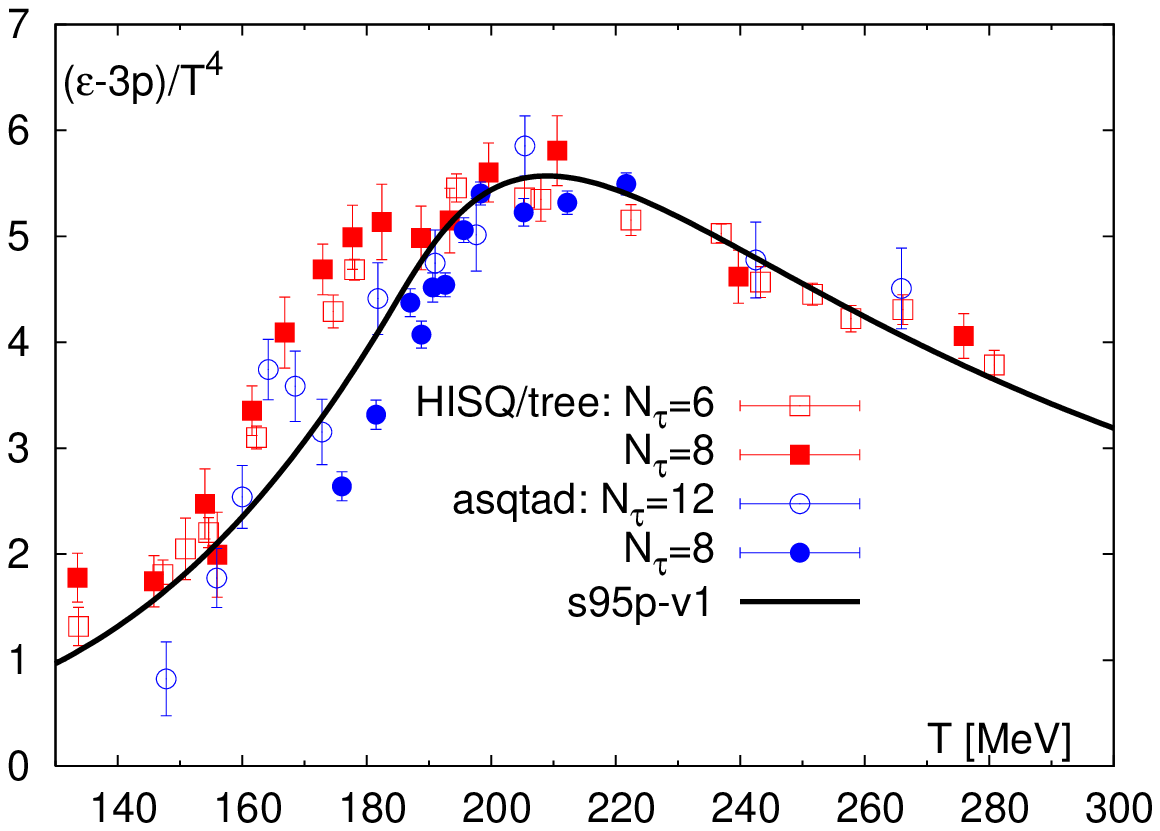}
\parbox[b]{0.48\textwidth}{
\caption{
$u-$ and $s-$quark and $u-$ and $d-$quark correlations 
rescaled by the corresponding fluctuations with $r_1$ scale
for HISQ/tree.
}
\label{fig:qns_us}
}
\hfill
\parbox[b]{0.48\textwidth}{
\caption{
The trace anomaly for the asqtad and HISQ/tree action,
$m_l=m_s/20$. ``s95p-v1'' parametrization is explained in
the text.
}
\label{fig:EoS}
}
\end{center}
\end{figure}

Correlations of different quark numbers are also a convenient way to study
the deconfinement aspects of the transition. At high temperatures such
correlations are expected to be very small, while at low temperatures
hadronic degrees of freedom give naturally rise to such correlations.
In Fig.~\ref{fig:qns_us} we present our results for the light-strange and
light-light quark number correlations. We divide them, correspondingly,
by the strange and light quark number susceptibility in attempt to get rid
of the trivial mass effects. For the strange-light quark number correlations
we see a good agreement with the HRG model, while the discrepancies with HRG
predictions for $\chi_{ud}/\chi_l$ are likely due to cut-off effects. 

Comparing Figs.~\ref{fig:qns_s}--\ref{fig:qns_us} 
we can conclude that quantities, where pions
do not contribute, can be reproduced quite well on the lattice,
while those, for which pions give dominant contribution, still
demonstrate noticeable cut-off effects, that are not completely
removed by switching to a hadronic ($f_K$) scale (at least,
on the lattices typically in use, $N_\tau=6-12$).

\section{Update on the trace anomaly}

The deconfinement transition is also often defined as the rapid
rise in the energy density associated with liberation of new
degrees of freedom. The energy density and other
thermodynamic quantities are usually obtained by integrating
the trace anomaly $(\epsilon-3p)/T^4$. This quantity is
under extensive investigation on the lattice. We present
an update of the results reported in \cite{Jamaica} in 
Fig.~\ref{fig:EoS}. The solid line is a parametrization
of $(\epsilon-3p)/T^4$ derived in Ref.~\cite{PasiPeter} that combines
HRG result at low $T$ ( $T<160$ MeV) with the lattice data \cite{hoteos}  at high $T$ 
($T>220$ MeV ). From the figure we conclude that the lattice results overshoot
the HRG prediction for $T>160$MeV since any cut-off effects could only decrease
$(\epsilon-3p)/T^4$ at low $T$.

\section{Conclusions}
In this contribution we studied the chiral transition in 2+1 flavor QCD
close to the physical point. Using the $O(N)$ scaling we determined the
chiral transition temperature in the continuum limit at the physical
light quark mass  to be $154(9)$~MeV. We also studied the
deconfinement phenomenon in terms of the renormalized
Polyakov loop as well as in terms of fluctuations and correlations of
conserved charges. We concluded that it is difficult to study the deconfinement
aspects of the transition in terms of the Polyakov loop, while correlations and fluctuations
of conserved charges are much better suited for this purpose. If $f_K$ is used
to set the scale we see a good agreement with the HRG model predictions for fluctuations
and correlations that do not involve pions in their HRG expansion.
For the quantities like $\chi_l$ and $\chi_{ud}$, where the pion contribution
is significant, cut-off effects are still too large for a meaningful comparison
with the HRG model.

\end{document}